\documentclass[11pt]{article}
\usepackage{amsmath,amssymb,color,graphics,epsfig,cite}

\usepackage{amsfonts}

\newcommand{\hoch}[1]{$\, ^{#1}$}

\makeatletter
\@addtoreset{equation}{section}
\makeatother

\newcommand{\be}{\begin{equation}}
\newcommand{\ee}{\end{equation}}
\newcommand{\bea}{\setlength\arraycolsep{2pt} \begin{eqnarray}}
\newcommand{\eea}{\end{eqnarray}}
\newcommand{\nn}{\nonumber}

\def\crampest{\medmuskip = 1mu plus 1mu minus 1mu}
\def\uncramp{\medmuskip = 4mu plus 2mu minus 4mu}

\def\ft#1#2{{\textstyle{\frac{\scriptstyle #1}{\scriptstyle #2} } }}
\def\fft#1#2{{\frac{#1}{#2}}}

\def\0{{\sst{(0)}}}
\def\1{{\sst{(1)}}}
\def\2{{\sst{(2)}}}
\def\3{{\sst{(3)}}}
\def\4{{\sst{(4)}}}
\def\5{{\sst{(5)}}}
\def\6{{\sst{(6)}}}
\def\7{{\sst{(7)}}}
\def\8{{\sst{(8)}}}
\def\sst#1{{\scriptscriptstyle #1}}

\def\del{{\partial}}

\begin{document}

\begin{flushright}
\hfill{MI-TH-1533}

\end{flushright}

\begin{center}
{\Large {\bf Black Hole Entropy and Viscosity Bound\\
 in Horndeski Gravity}}

\vspace{15pt}
{\bf Xing-Hui Feng\hoch{1}, Hai-Shan Liu\hoch{2,3}, H. L\"u\hoch{1} and C.N. Pope\hoch{3,4}}

\vspace{10pt}

\hoch{1}{\it Center for Advanced Quantum Studies, Department of Physics, \\
Beijing Normal University, Beijing 100875, China}

\vspace{10pt}

\hoch{2} {\it Institute for Advanced Physics \& Mathematics,\\
Zhejiang University of Technology, Hangzhou 310023, China}

\vspace{10pt}

\hoch{3} {\it George P. \& Cynthia Woods Mitchell  Institute
for Fundamental Physics and Astronomy,\\
Texas A\&M University, College Station, TX 77843, USA}

\vspace{10pt}

\hoch{4}{\it DAMTP, Centre for Mathematical Sciences,
 Cambridge University,\\  Wilberforce Road, Cambridge CB3 OWA, UK}

\vspace{20pt}

\underline{ABSTRACT}
\end{center}

Horndeski gravities are theories of gravity coupled to a scalar field,
in which the action contains an additional non-minimal quadratic
coupling of the
scalar, through its first derivative, to the Einstein tensor or the
analogous higher-derivative tensors coming from the variation of
Gauss-Bonnet or Lovelock terms. In this paper we study the thermodynamics of
the static black hole solutions in $n$ dimensions,
in the simplest case of a Horndeski
coupling to the Einstein tensor.  We apply the Wald formalism to
calculate the entropy of the black holes, and show that there is
an additional contribution over and above those that come from the
standard Wald entropy formula.  The extra contribution can be attributed
to unusual features in the behaviour of the scalar field.  We also show
that a conventional regularisation to calculate the Euclidean action
leads to an expression for the entropy that disagrees with the Wald
results.  This seems likely to be due to ambiguities in the subtraction
procedure.   We also calculate the viscosity in the dual CFT, and show that
the viscosity/entropy ratio can violate the $\eta/S\ge 1/(4\pi)$ bound for
appropriate choices of the parameters.

\vfill {\footnotesize xhfengp@mail.bnu.edu.cn \ \ \  hsliu.zju@gmail.com \ \ \ mrhonglu@gmail.com\ \ \
pope@physics.tamu.edu}

\thispagestyle{empty}

\pagebreak

\tableofcontents
\addtocontents{toc}{\protect\setcounter{tocdepth}{2}}



\section{Introduction}

In the dictionary of gravity/gauge duality mappings
in the AdS/CFT correspondence
\cite{adscft1,adscft2,adscft3}, perturbations of the metric are related to
the energy-momentum tensor of the field theory in the boundary of the
AdS spacetime
\cite{adscft2,adscft3,adscft4}. In this picture, an AdS planar black hole is
the gravitational dual of a certain ideal fluid.  A widely valid
relation between the shear viscosity and the entropy density was established,
namely \cite{Policastro:2001yc,sonsta,KSS,KSS0}
\be
\fft{\eta}{S}=\fft{1}{4\pi}\,.\label{visent0}
\ee
One way to understand this ratio is that it can be shown that the
viscosity is proportional to the cross-section of the black hole for
low-frequency
massless scalar fields \cite{KSS0}. Alternatively, the shear viscosity is
determined by the effective coupling constant of the transverse graviton
on the horizon, by employing the membrane paradigm \cite{Iqbal:2008by}.
(This was confirmed by using the Kubo formula in
\cite{Cai:2008ph,Cai:2009zv}.) In \cite{Brustein:2007jj}, it was shown
that the black hole entropy is determined by the effective Newtonian
coupling
at the horizon, and that it is thus not surprising that the ratio of
the shear viscosity to the entropy density is universal, in the sense that
the dependence of the quantities on the horizon is canceled.  Recently,
it was established that the relation (\ref{visent0}) of the boundary
theory is dual to a generalised Smarr relation obeyed by the bulk AdS
planar black holes, thereby providing a new understanding of its
universality, and its connection to the black hole thermodynamics
\cite{Liu:2015tqa}.  There have been a number of papers in literature
establishing the universality of the ratio
(\ref{visent0}) \cite{Buchel:2003tz,Buchel:2004qq,Benincasa:2006fu,Landsteiner:2007bd}. (See \cite{Cremonini:2011iq} for a review.)

   The viscosity/entropy ratio (\ref{visent0}) can, however, be violated
when the bulk gravity theory is extended by the addition of
higher-order curvature terms \cite{Kats:2007mq,shenker}.\footnote{We shall
not be concerned in this paper with other types of violation, due to
the breaking of local rotational symmetry; see, for example,
\cite{Natsuume:2010ky,Erdmenger:2010xm,Ovdat:2014ipa,Ge:2014aza}.}
(See also, for further examples,
\cite{Shu:2009ax,deBoer:2009pn,Camanho:2009vw}.)

   This leads us to one of
the motivations for this paper, which is to investigate whether one
can violate the ratio (\ref{visent0}) without introducing higher-order
curvature terms in the bulk theory.   In a typical theory of
Einstein gravity, matter fields couple to gravity minimally through
the metric. A scalar field can also couple to gravity non-minimally,
such as in Brans-Dicke theory \cite{Brans:1961sx}, where the effective
Newton constant varies in spacetime.  However, it was established in
\cite{Liu:2015tqa} that the ratio (\ref{visent0}) holds in general
in such a theory.  Scalar fields can, however, also couple non-minimally
to gravity in other ways.  In particular, their derivatives can couple to
the curvature tensor. Horndeski considered a wide class of such gravity/scalar
theories in the early seventies \cite{Horndeski:1974wa}, focusing his
attention on cases where the field equations, both for gravity and the
scalar field, involve no higher than second derivatives.  The Horndeski
theories were rediscovered recently in studies of the covariantisation
of Galileon theories \cite{Nicolis:2008in}.

   The Horndeski terms take the
form
\be
H^{(k)} = E_{\mu\nu}^{(k)} \del^\mu \chi \del^\nu\chi\,,
\ee
where the $E^{(k)}$ tensors are ``energy-momentum tensors'' associated
with the Euler integrands of various order, namely
\be
E^{(k)\,\nu}{}_\mu \equiv \delta^{\nu\rho_1\cdots\rho_{2k}}_{\mu\sigma_1\cdots
\sigma_{2k}}\, R^{\sigma_1\sigma_2}{}_{\rho_1\rho_2}\,\cdots\,
R^{\sigma_{2k-1}\sigma_{2k}}{}_{\rho_{2k-1}\rho_{2k}}\,.
\ee
The $H^{(k)}$ terms are analogous to Euler integrands, in that they
have the property
that each field carries no more than a single derivative and hence
the linearized equations of motion involve at most second derivatives.
Thus although the theory involves higher-order derivatives, it contains
no linear ghost excitations.  In this paper, we shall consider Einstein
gravity with a cosmological constant, together with just the two lowest-order
Horndeski terms, namely
\be
H^{(0)}= g_{\mu\nu} \del^\mu\chi \del^\mu \chi\,,
\qquad H^{(1)} =-4 G_{\mu\nu} \del^{\mu}\chi \del^\nu\chi
\,,\label{horndeski01}
\ee
where $G_{\mu\nu}$ is the Einstein tensor.  We find that although the
theory contains the curvature tensor only linearly,
the viscosity/entropy ratio (\ref{visent0}) no longer holds.

It is worth commenting that the viscosity can be computed by
standard procedures using the AdS/CFT correspondence, involving
the straightforward technique of studying linearised perturbations
around the background bulk solution.  The calculation of the
viscosity/entropy ratio then
hinges upon the proper definition of the entropy of the black hole.
Since Hawking established the thermal radiation of a black hole
\cite{Hawking:1974rv,Hawking:1974sw}, there has been no ambiguity in
establishing the black hole entropy in a generally-covariant theory.
In particular, in Einstein gravity minimally coupled to matter, the
entropy is given by one quarter of the area of the horizon. This area law
has been generalized to the Wald entropy formula when more complicated
couplings
or higher-order curvature terms are involved, namely \cite{wald1,wald2}
\be
S_{\sst W}=-\fft{1}{8} \int_+ d^{n-2} x \sqrt{h}\,
\fft{\partial L}{\partial R^{abcd}}
\epsilon^{ab} \epsilon^{cd}\,.\label{waldentropy}
\ee
where $L$ is defined by the action $I=\int d^n x \sqrt{-g} L$.
Applying this formula to static black holes with spherical, toric or
hyperbolic isometries, the Horndeski terms (\ref{horndeski01}) do not
contribute to the Wald entropy $S_{\sst W}$, and hence one might expect that
the entropy would still be just one quarter of the horizon area.
However, we find that this is in fact not the case.
By examining the Wald procedure \cite{wald1,wald2} in detail, we find that
in a theory such as Horndeski gravity there is an additional contribution
to the entropy that is not encompassed by the usual Wald
formula (\ref{waldentropy}).  It arises because the
derivative of the scalar field diverges
on the horizon in the black-hole solutions (although there is no physical
divergence, since all invariants, such as
$g^{\mu\nu}\, \del_\mu\chi\, \del_\nu\chi$, remain finite).

The paper is organised as follows.  In section 2 we introduce the
Horndeski theory that we shall be considering, and we review the
static black hole solutions.  These are known for all the cases of
spherical, toroidal and hyperbolic horizon geometries.  Our focus
will be on the spherical and the toroidal horizons.  We also include
a demonstration of the uniqueness of the known static solutions. In
section 3 we address the problem of calculating the entropy, and also
the mass, of the static black holes.  We begin by calculating the entropy
using the standard Wald formula (\ref{waldentropy}), and then we
consider the application of the Wald formalism in more detail, showing that
there is another contribution to the entropy that is not captured
by (\ref{waldentropy}).  We show that in the case of the planar black holes
(with toroidal horizons), the entropy expression we obtain is consistent
with the computation of the Noether charge associated with a scaling
symmetry of the black holes.  We also consider the calculation of
the Euclidean action, showing that, at least when following a
naive regularisation procedure, this yields yet another result for the
entropy, and the mass, that disagrees with those from the Wald formalism.
In section 4 we calculate the shear viscosity in the dual boundary theory
using the AdS/CFT correspondence, and hence we obtain an expression
for the viscosity/entropy ratio.  This is different from $1/(4\pi)$ on
account of the Horndeski term, and we show that
for an appropriate choice of the
parameters it can violate the $\eta/S\ge 1/(4\pi)$ bound.  The paper
ends with conclusions in section 5.

\section{Black Holes in Horndeski Gravity}

\subsection{The theory}
\label{thetheory}

As we have discussed in the introduction, Horndeski gravity represents a
class of higher-derivative theories involving gravity with a non-minimally
coupled scalar.  The couplings differ from those in the Brans-Dicke
theory, since in the Horndeski theories the scalar couples through its
derivative to the curvature tensors. We shall focus on the Horndeski
theory whose Lagrangian
involves at most only linear curvature terms.
As we shall show, the viscosity/entropy
ratio (\ref{visent0}) can be violated even in such a theory.
The action is given by
\be
I = \fft{1}{16\pi}\int d^nx\sqrt{-g}\, L\,,\qquad
L = \kappa(R-2\Lambda)-\ft{1}{2}(\alpha g_{\mu\nu}-\gamma G_{\mu\nu})
\del^\mu\chi\, \del^\nu\chi\,,\label{action}
\ee
where $\kappa$,  $\alpha$ and  $\gamma$ are coupling constants,
and $G_{\mu\nu}\equiv R_{\mu\nu} - \ft12 R\,g_{\mu\nu}$ is the Einstein
tensor.  Note that the theory is invariant under a constant shift of
$\chi$.  In a typical gravity theory with a scalar field, such as
Brans-Dicke theory, one can define different metric frames by means
of conformal scalings using the scalar field.  However, for the
Horndeski theory (\ref{action}), this would lead to the breaking of the
manifest constant shift symmetry of the scalar,  and hence it would
not be a  natural field redefinition to make here.

The variation of the action (\ref{action}) gives rise to
\be
\delta I = \fft1{16\pi}\,
  \int d^n x \sqrt{-g} (E_{\mu\nu} \delta g^{\mu\nu} + E \delta \chi +
\nabla_\mu J^\mu)\,.\label{variation}
\ee
where
\bea
E_{\mu\nu} &=& \kappa (G_{\mu\nu} +\Lambda g_{\mu\nu}) -
\ft12\alpha \Big(\partial_\mu \chi \partial_\nu \chi - \ft12 g_{\mu\nu} (\partial\chi)^2\Big)-\ft12\gamma \Big(\ft12\partial_\mu\chi \partial_\nu \chi R - 2\partial_\rho
\chi\, \partial_{(\mu}\chi\, R_{\nu)}{}^\rho \cr
&&- \partial_\rho\chi\partial_\sigma\chi\, R_{\mu}{}^\rho{}_\nu{}^\sigma -
(\nabla_\mu\nabla^\rho\chi)(\nabla_\nu\nabla_\rho\chi)+(\nabla_\mu\nabla_\nu\chi)
\Box\chi + \ft12 G_{\mu\nu} (\partial\chi)^2\cr
&&-g_{\mu\nu}\big[-\ft12(\nabla^\rho\nabla^\sigma\chi)
(\nabla_\rho\nabla_\sigma\chi) + \ft12(\Box\chi)^2 -
  \partial_\rho\chi\partial_\sigma\chi\,R^{\rho\sigma}\big]\Big)\,,\cr
E &=&\nabla_\mu \big( (\alpha g^{\mu\nu} - \gamma G^{\mu\nu}) \nabla_\nu\chi\big)\,.
\eea
The total derivative term in (\ref{variation}) plays no role in the
equations of motion
\be
E_{\mu\nu}=0\,,\qquad E=0\,.
\ee
However, it does play an important role in the Wald formalism, which we
shall present in section \ref{waldformalism}.

\subsection{Static black hole solutions}

We now consider static black holes, with the ansatz
\be
ds_n^2 = - h(r) dt^2 + \fft{dr^2}{f(r)} + r^2 d\Omega_{n-2,\epsilon}^2\,,\qquad
\chi=\chi(r)\,,\label{bhansatz}
\ee
where $d\Omega_{n-2,\epsilon}^2$ with $\epsilon=1,0,-1$ is the metric
for the unit $S^{n-2}$, the $n$-torus or the unit hyperbolic $n$-space.
It is convenient to take $d\Omega_{n-2,\epsilon}^2=\bar g_{ij} dy^i dy^j$
for general values
of $\epsilon$ to be the metric of constant curvature such that its
Ricci tensor is given by $\bar R_{ij}= (n-3)\, \epsilon\, \bar g_{ij}$.  We
may, for example, take $d\Omega_{n-2,\epsilon}^2$ to be given by
\be
d\Omega_{n-2,\epsilon}^2 = \fft{du^2}{1-\epsilon u^2} +
    u^2\, d\Omega_{n-3}^2\,,
\ee
where $d\Omega_{n-3}^2$ is the metric of the unit $(n-3)$-sphere.

It is clear from the equations of motion that $\chi=\chi_0$ (constant)
is a solution, in which case, the Horndeski gravity reduces to
Einstein gravity with a cosmological constant $\Lambda_0$.  It follows
that the Schwarzschild-AdS black hole is a solution of the theory.
We shall regard this solution as being ``trivial,'' in the sense
of not yielding anything new.
In addition, a one-parameter family of black hole solutions for which
the scalar field is not a constant was constructed in \cite{aco}.
(See also,\cite{Rinaldi:2012vy,Babichev:2013cya}.)
In this section, we would like to prove that these are the only black hole
solutions from the ansatz (\ref{bhansatz}) in which the scalar
is $r$-dependent.  First, we  review the construction in \cite{aco}.

   The scalar equation of motion $E=0$ yields
\be
\Big(r^{n-4} \sqrt{\fft{f}{h}}
\Big( \gamma \big( (n-2) r f h' + (n-2)(n-3) (f-\epsilon) h\big)
    -2\alpha r^2 h \Big)\chi'\Big)'=0\,.\label{scalareom}
\ee
There are two more equations that follow from $E_{\mu\nu}=0$:
\bea
&&4\kappa \Big((n-2)r f' + (n-2)(n-3) (f-\epsilon) +
2\Lambda_0 r^2\Big) + 2\alpha r^2 f\chi'^2\cr
&&\qquad+\gamma (n-2) \Big(4r f \chi'' + \big( 3r f' +
   (n-3)(f+\epsilon)\big)\chi' \Big)f \chi'
=0\,,\cr
&&4\kappa \Big((n-2) r f h' + (n-2)(n-3) h (f-\epsilon) + 2\Lambda_0
    r^2 h\Big) -2 \alpha r^2 f h \chi'^2\cr
&&\qquad + \gamma (n-2) \Big( 3 r f h' + (n-3) (3 f - \epsilon)h\Big) f \chi'^2
=0\,.
\eea

   In \cite{aco}, a class of black hole solution was obtained by
solving (\ref{scalareom}) by taking
\be
\gamma \big( (n-2) r f h' + (n-2)(n-3) (f-\epsilon) h\big) -2\alpha r^2 h=0\,.
\label{specialeom}
\ee
(In other words, the integration constant in the first integral of
(\ref{scalareom}) was taken to be zero, and $\chi'$ was allowed to
be non-zero, thus implying that its co-factor, given in (\ref{specialeom}),
must be equal to zero.)
This leads to the solution
\bea
h &=& -\fft{\mu}{r^{n-3}} + \fft{8\kappa
[ g^2r^2 (2\kappa + \beta\gamma)+2\epsilon\kappa]}{(4\kappa + \beta\gamma)^2}
\cr
&&+\fft{(n-1)^2\,
\beta^2\gamma^2 g^4 r^4}{\epsilon\, (n+1)(n-3)(4\kappa + \beta\gamma)^2}
\,\,{}_2F_1\Big[1, \ft12(n+1); \ft12(n+3);
      -\fft{n-1}{(n-3)\epsilon} g^2 r^2\Big]\,,\cr
f &=& \fft{(4\kappa + \beta\gamma)^2 \big[(n-1)g^2r^2 +
     (n-3)\epsilon\big]^2}{
\big[(n-1)(4\kappa + \beta\gamma) g^2 r^2 +
   4(n-3)\epsilon\kappa\big]^2}\,h\,,\qquad
\chi'^2 = \fft{\beta}{f}\,
  \Big[1 + \fft{(n-3)\epsilon}{(n-1) g^2 r^2}\Big]^{-1}\,,
\label{gensol}
\eea
which is valid for all values of $\epsilon$.
In presenting the solution, we have introduced two parameters
$(g,\beta)$ in place of the original parameters $(\alpha, \Lambda_0)$
in the Lagrangian, with
\be
\alpha=\ft12 (n-1)(n-2) g^2 \gamma\,,\qquad
\Lambda_0=-\ft12 (n-1)(n-2)g^2 \Big(1 + \fft{\beta\gamma}{2\kappa}\Big)\,.
\label{alphabeta}
\ee
Note that the solution contains only one integration constant, $\mu$.
All other parameters are those of the theory itself.  Note also that since
the dimension $n$ is an integer, the hypergeometric function reduces to
polynomials with an $\arctan$ function in even dimensions, and with a
$\log$ function in odd dimensions. To be explicit,  we have
\bea
&&\qquad\qquad\qquad n =\hbox{even}:\\
\!\!\! &&\!\!\! {}_2F_1[1, \ft12(n+1); \ft12(n+3);-x]=
  \fft{(-1)^{n/2}(n+1)}{x^{n/2}}\,
 \Big\{\fft{\arctan\sqrt{x}}{\sqrt{x}}-
      \Big[\fft{\arctan\sqrt{x}}{\sqrt{x}}\Big]_{\ft{n}{2}-1}\Big\}\,,\nn\\
&&\qquad\qquad\qquad n = \hbox{odd}:\\
\!\!\! &&\!\!\! {}_2F_1[1, \ft12(n+1); \ft12(n+3);-x]=
\fft{(-1)^{\ft{n-1}{2}}\, (n+1)}{2 x^{\ft{n-1}{2}}}\,
\Big\{ \fft{\log(1+x)}{x} - \Big[\fft{\log(1+x)}{x}\Big]_{\ft{n-3}{2}}\Big\}\,,
\nn
\eea
where we use the notation $[F(x)]_m$ to denote the truncated
power series expansion of $F(x)$ around $x=0$, in which only the terms up
to and including $x^m$ are retained. Thus
\be
\Big[\fft{\arctan\sqrt{x}}{\sqrt{x}}\Big]_{\ft{n}{2}-1}=
\sum_{p=0}^{n/2-1}\, \fft{(-x)^p}{2p+1}\,,\qquad\qquad
\Big[\fft{\log(1+x)}{x}\Big]_{\ft{n-3}{2}} = \sum_{p=0}^{\ft{n-3}{2}}\,
\fft{(-x)^p}{p+1}\,,
\ee
for $n$ even and $n$ odd, respectively.

For static solutions of this kind, it is in fact always sufficient to
construct
the solution with $\epsilon=1$.  The solutions for all other values
of $\epsilon$, which we presented above,
can then be obtained from the $\epsilon=1$ solution
by means of the rescalings
\be
r\longrightarrow \fft{r}{\sqrt{\epsilon}}\,,\qquad
t\longrightarrow \sqrt{\epsilon}\, t\,,\qquad
d\Omega_{n-2}^2 \longrightarrow \epsilon\, d\Omega_{n-2,\epsilon}^2\,,
\qquad
\mu\longrightarrow \epsilon^{-(n-1)/2}\, \mu
\ee
 From now on, we shall present results for the two specific
cases $\epsilon=0$ and $\epsilon=1$.

\medskip
\noindent{\underline{\bf $\epsilon=0$ solution}:}
\medskip

When $\epsilon=0$, the solution reduces to the very simple form
\be
h=f=g^2 r^2 - \fft{\mu}{r^{n-3}}\,,
\qquad \chi'^2 = \fft{\beta}{f}\,. \label{planarso}
\ee
Note that in this $\epsilon=0$ case, $\chi$ can be solved for explicitly,
giving
\be
\chi=\fft{2\sqrt{\beta}}{(n-1)g} \log\big( \sqrt{(gr)^{n-1}} +
\sqrt{(gr)^{n-1} - \mu g^{n-3}}\big) + \chi_0\,.\label{phiso1}
\ee
Thus the $\epsilon=0$ solution describes an AdS planar black hole,
with the
requirements that $\mu> 0$ and $\beta\ge 0$. The horizon radius $r=r_0$ is
given by $\mu=g^2 r_0^{n-1}$.  The Hawking temperature is
given by
\be
T=\fft{(n-1)g^2}{4\pi} r_0\,.
\ee

\medskip
\noindent{\underline{\bf $\epsilon=1$ solution}:}
\medskip

For $\epsilon=1$, the solution describes  a spherically-symmetric and
static black hole.  In a large-$r$ expansion, if $n$ is even
the functions $h$ and $f$ have the asymptotic forms
\bea
h &=& g^2r^2 - \fft{\mu}{r^{n-3}} + \sum_{k=0} \fft{c_k}{r^{2k}}=
g^2 r^2 + \fft{4\kappa - \beta\gamma}{4\kappa +
\beta\gamma}\epsilon + \cdots\,,\cr
f &=& g^2r^2 - \fft{\mu}{r^{n-3}} + \sum_{k=0} \fft{d_k}{r^{2k}}=
g^2 r^2 + \fft{4(n-1)\kappa + (n-5)\beta\gamma}{
    (n-1) (4\kappa + \beta\gamma)}\epsilon + \cdots\,,\label{larger}
\eea
where $(c_k,d_k)$ are constants, which are functions of the
parameters $(\kappa,g,\beta)$
but independent of $\mu$.  If $n$ is odd, then for $k=(n-3)/2$, the quantity
$c_k$
has an additional term proportional to $\log r$.  This amounts to a
logarithmically diverging addition to the mass coefficient $\mu$ at
order $1/r^{n-3}$.  This in turn implies
that $d_k$ has additional $\log r$ terms for all $k\ge (n-3)/2$.  Note that
all the $(c_k,d_k)$ vanish for $\epsilon=0$.

The metric is asymptotic locally to AdS spacetime,  and it cannot become
pure AdS spacetime, regardless of the choice of the parameter $\mu$.
To see that the solution describes a black hole, we note that $h$ is
positive as $r$ goes to infinity, but becomes of order $-\mu/r^{n-3}$ as
$r\rightarrow 0$, where there is a spacetime curvature singularity.  Thus
when $\mu>0$, there must exist some intermediate value of $r$,
be an event horizon $r=r_0$, for which
\be
h(r_0)=0=f(r_0)\,.\label{horizoneq}
\ee
This implies that the parameter $\mu$ can be expressed in terms of the
horizon radius $r_0$ in this $\epsilon=1$ case as
\bea
\mu &=& \fft{8\kappa r_0^{n-3}}{(4\kappa + \beta\gamma)^2}\Big(
2\kappa + (2\kappa + \beta\gamma)g^2 r_0^2\cr
&& + \fft{(n-1)^2 \beta^2\gamma^2 g^4 r_0^4}{8\kappa (n-1)(n-3)}
\,{}_2F_1[1, \ft12(n+1); \ft12(n+3); -\ft{n-1}{(n-3)} g^2 r_0^2]
\Big)\,.\label{mueq}
\eea
Note that this relation between $\mu$ and $r_0$ is far more complicated
than the simple expression $\mu=g^2\, r_0^{n-1}$ that holds in
the $\epsilon=0$ case.
The temperature of the $\epsilon=1$ black hole is given by
\be
T=\fft{\sqrt{h'(r_0) f'(r_0)}}{4\pi} = \fft{(n-1)g^2}{4\pi} r_0+
\fft{(n-3) \kappa}
{\pi (4\kappa + \beta\gamma)r_0}\,.\label{temp2}
\ee
Note that if we set $\mu=0$, then the solution has no event horizon, and
near $r=0$ the functions $h$, $f$  and $\chi$ have the forms
\bea
h &=& \fft{16\kappa^2}{(4\kappa + \beta\gamma)^2} \Big(1 + \fft{(2\kappa + \beta\gamma) g^2 r^2}{2\kappa} + \cdots\Big)\,,\cr
f &=& 1 + \fft{((n-3)\kappa -\beta\gamma) g^2 r^2}{(n-3)\kappa} + \cdots\,,\cr
\chi &=& \chi_0 + \fft{(n-1)\beta}{2(n-3)} gr^2 + \cdots\,.
\eea
Thus the $\mu=0$ solution is a smooth spherically-symmetric soliton,
without any free parameters, that is asymptotic locally to AdS spacetime.
There also exists a solution for $\epsilon=1$ in the limit of
$4\kappa + \beta \gamma=0$, but it does not describe a black hole.

\subsection{Uniqueness of the Horndeski black hole solutions}

We shall leave the discussion of the mass and entropy of the black holes
to the next section.  To close this section, we shall
show that the solutions discussed above are in fact the only black holes
with non-constant  $\chi$ that are contained within the ansatz (\ref{bhansatz})
in the theory. To show this, we return to the equation of motion
 (\ref{scalareom}) for the scalar field.  One can immediately write down
the first integral
\be
\chi'=\fft{q\, r^{4-n} \sqrt{h/f}}{\gamma \big( (n-2) r f h' +
(n-2)(n-3) (f-\epsilon) h\big) -2\alpha r^2 h}\,,\label{phiprime}
\ee
where $q$ is an integration constant.  The solutions we discussed above
were obtained by taking $q=0$.  It was possible to find such solutions
with $\chi'\ne0$ by imposing the relation (\ref{specialeom}), which in fact
rendered the scalar equation of motion (\ref{scalareom}) trivial.  If instead
we take the integration constant $q$ to be non-zero, then $\chi'$ is
now determined by (\ref{phiprime}).

    If a solution with $q\ne0$ is to describe describe a black hole,
there must be an event horizon at some radius $r=r_0$.  The functions
$h$ and $f$ near
the horizon will have Taylor expansions of the form
\be
f=f_1(r-r_0) + f_2 (r-r_0)^2 + \cdots\,,\qquad
h=h_1 (r-r_0) + h_2 (r-r_0)^2 + \cdots\,.\label{hfexpan}
\ee
It follows from (\ref{phiprime}) that $\chi'$ near the horizon has the
expansion
\be
\chi' = \fft{\tilde \chi_{-1}}{r-r_0} + \tilde \chi_0 +
\tilde \chi_1 (r-r_0) + \cdots\,.
\ee
Substituting these expansions into the other equations of motion, we find
that no such solutions can exist.  In other words, the assumption
that there exists a horizon, near which the expansions (\ref{hfexpan})
would hold,  is inconsistent with the equations of motion when
$q\ne0$..  In order to have a solution
with  a horizon, we must therefore set $q=0$, which then reduces to
the previous case discussed above.  However,
as mentioned already, in order for this solution not to be trivial, i.e. for
$\chi'$ to be non-vanishing,
we must then also impose the condition (\ref{phiprime}). This leads
the to the black hole solution (\ref{gensol}).

  In the near-horizon region,
the function $\chi$ in the black-hole solutions (\ref{gensol})
has an expansion of the form
\be
\chi=\tilde \chi_0 + \tilde \chi_1 (r-r_0)^{\fft12} +
\tilde \chi_2 (r-r_0)^{\fft32} + \cdots\,.\label{phiexpan}
\ee
Substituting back into the equations of motion, we find that all the
coefficients in the expansions can be expressed in terms of two
parameters, $h_1$ and $r_0$. For example,
\be
f_1 = \fft{(n-2)(n-3)\gamma \epsilon + 2\alpha r_0^2}{(n-2)\gamma r_0}\,,
\qquad
\chi_1 = \fft{2\sqrt{(n-1)\beta} g r_0^{\fft32}}{(n-1)g^2 r_0^2 +
    (n-3)\epsilon}\,,\qquad
\cdots\,.
\ee
Thus the solution has three integration constants
$(\tilde \chi_0, h_1,r_0)$.  However, the parameters $(\tilde \chi_0, h_1)$
are trivial.  It follows that the only non-trivial parameter is $r_0$,
which is determined by $\mu$ in the final solution.

Finally we would like to emphasize again that $\beta$ is not an
integration constant, but a parameter of the theory.  For $\beta\ne 0$,
there are two black holes, but each associated with a different vacuum.
When $\beta=0$, there is only the Schwarzschild-AdS black hole solution
in the theory.

\section{Black Hole Entropy and Thermodynamics}

In the previous section, we reviewed the Horndeski gravity theory, and
its static black hole solutions.  We identified the horizon and computed
the temperature of these black holes.
In this section, we consider various possible methods for calculating
their entropy.  It turns out that different well-established methods yield
different answers.  A correct answer of the entropy is important for studying
the black hole thermodynamics, and it is paramount for determining
the $\eta/S$ ratio, as we discussed in the introduction.

\subsection{Wald entropy formula}\label{waldentropysec}

First let us consider the well-known Wald entropy formula (\ref{waldentropy}).
It is straightforward to see that for the Horndeski Lagrangian $L$ given
in (\ref{action}), one has
\crampest
\bea
T^{\mu\nu\rho\sigma} &\equiv& \fft{\del L}{\del R_{\mu\nu\rho\sigma}}=
\ft12\kappa\, (g^{\mu\rho}\, g^{\nu\sigma} - g^{\nu\rho}\, g^{\mu\sigma})
\label{dLdR}\\
&&+
\ft18 \gamma\, [g^{\mu\rho}\, \chi^\nu\, \chi^\sigma -
  g^{\nu\rho}\,\chi^\mu\, \chi^\sigma + g^{\nu\sigma}\,\chi^\mu\, \chi^\rho
- g^{\mu\sigma}\, \chi^\nu\, \chi^\rho -
(g^{\mu\rho}\, g^{\nu\sigma} - g^{\nu\rho}\, g^{\mu\sigma})\, \chi^\lambda\,
\chi_\lambda ]\,,\nn
\eea
\uncramp
where we have defined $\chi_\mu=\del_\mu\chi$.
For the static black holes in the
Horndeski theory, described in section 2, we find from (\ref{dLdR}) that
the Wald entropy formula (\ref{waldentropy})
for the entropy gives the same result as in standard Einstein gravity,
namely one quarter of the area of the event horizon,
\be
S_{\sst W}=\ft14\kappa r_0^{n-2} \, \omega_{n-2}\,,\label{S0}
\ee
where
$\omega_{n-2}$ is the volume of a unit $S^{n-2}$ in the $\epsilon=1$ case.
For $\epsilon=0$, corresponding to a toroidal horizon, the periods of
the circles forming the torus can be chosen arbitrarily, and we shall,
for convenience, then take $\omega_{n-2}=1$ in this paper, and so
correspondingly $S$ should then be
viewed as the entropy density.

  Since the static black hole solutions are characterised by only one
parameter (i.e. one integration constant),
it is guaranteed that one can obtain an expression for a
``thermodynamic mass'' by integrating
the first law of black hole thermodynamics\footnote{In a more general
situation where there are further intensive/extensive pairs of
thermodynamic variables contributing on the right-hand side of the
first law for multi-parameter solutions,
the integrability of the right-hand side can provide
a non-trivial check on the correctness of the thermodynamic
quantities.  No such consistency check arises in the case of a
one-parameter family of solutions, since all 1-forms are exact
in one dimension.}
\be
dM=TdS\,.\label{firstlaw}
\ee
If we use the expression (\ref{S0}) for the entropy, then from the result
for the Hawking temperature obtained in the previous section we
therefore find
\bea
\underline{\epsilon=0}:&&
M=\fft{\kappa (n-2)}{16\pi}\,\mu \,,\label{wfmass1}\\
\underline{\epsilon=1}:&&
M=\Big(\fft{\kappa(n-2)}{16\pi} g^2 r_0^{n-1}
+\fft{\kappa^2(n-2)}{4\pi (4\kappa + \beta \gamma)}
  r_0^{n-3}\Big)\, \omega_{n-2}\,.\label{wfmass2}
\eea
   Note that in the $\epsilon=0$ case it was straightforward to
express the mass in terms of the ``mass parameter'' $\mu$, because
of the simple relation $\mu= g^2\, r_0^{n-1}$ for these planar
black holes.  On the other hand, the relation between $\mu$ and
$r_0$ is much more complicated in the $\epsilon=1$ case, and is
given in (\ref{mueq}).  Thus when $\epsilon=1$ the expression
(\ref{wfmass2})
for $M$ would become a complicated transcendental function of the
mass parameter $\mu$.

    On the face of it, the mass
formula (\ref{wfmass1}) for the $\epsilon=0$ case looks
not unreasonable.   In fact the
thermodynamical quantities satisfy also the expected generalised
Smarr relation
\be
M=\fft{n-2}{n-1} \, T S_{\sst W}\,.\label{gsmarr}
\ee
However, for the $\epsilon=1$ case, the mass formula (\ref{wfmass2})
looks less reasonable.  As mentioned above, it would
be a complicated transcendental function of the ``mass parameter''
$\mu$.  Whilst this fact, of itself, does not conclusively show that
it must be incorrect, it does perhaps raise doubts
about its likely  validity, since it would be a very unusual kind
of relation that is not normally seen in other black hole solutions.
Furthermore,  if the $\epsilon=1$ mass formula
is called into question then this also raises questions about the
validity of the $\epsilon=0$ mass formula.

  In order to explore these issues in greater depth, we shall make a
more detailed investigation of the Wald procedure, in order to see whether
there are new subtleties that can arise in a theory such as that of
Horndeski.

\subsection{Wald formalism}
\label{waldformalism}

Wald has developed a procedure for deriving the first law of thermodynamics
by calculating the variation of a Hamiltonian derived from a conserved
Noether current.  The general procedure was presented in
\cite{wald1,wald2}. The Wald entropy formula (\ref{waldentropy}) is a
consequence of applying this procedure in rather generic
higher-derivative theories.  The Wald formalism has been used to study
the first law of thermodynamics for asymptotically-AdS black holes
in variety of theories, including
Einstein-scalar \cite{Liu:2013gja,Lu:2014maa},
Einstein-Proca \cite{Liu:2014tra}, Einstein-Yang-Mills \cite{Fan:2014ixa},
in gravities extended with quadratic-curvature
invariants \cite{Fan:2014ala}, and also for Lifshitz black holes
\cite{Liu:2014dva}. However, the rather unusual-looking results
that it led to for the mass of the $\epsilon=1$
black holes in section 3.1 raised the possibility that
the formula (\ref{waldentropy}) might not be valid for Horndeski gravity.
For this reason, we shall now study in detail the application of the
Wald formalism for the action (\ref{action}).

  A general variation of the fields in the
action (\ref{action}) was given in (\ref{variation}).  The surface
term $J^\mu$ is given by
\bea
J^\mu &=& 2 \fft{\partial L}{\partial R_{\rho\sigma\mu\nu}} \nabla_\sigma \delta g_{\rho \nu} -2 \nabla_\nu \fft{\partial L}{\partial R_{\rho\mu\nu\sigma} }\delta g_{\rho\sigma} + \fft{\partial L}{\partial(\nabla_\mu \chi)} \delta \chi\cr
&=&
\Big(\kappa J^\mu_g+\alpha J^\mu_\chi+\gamma(J^\mu_{gc}+J^\mu_{\chi c})\Big)\,,
\eea
with
\bea
J^\mu_g&=&g^{\mu\rho}g^{\nu\sigma}(\nabla_\sigma\delta g_{\nu\rho}-\nabla_\rho\delta g_{\nu\sigma})\,,\quad J^\mu_\chi=-g^{\mu\nu}\nabla_\nu\chi\delta\chi\,,\quad
J^\mu_{\chi c}=G^{\mu\nu}\nabla_\nu\chi\delta\chi\,,\cr
J^\mu_{gc}&=&-\ft{1}{4}(\nabla\chi)^2J^\mu_g+\ft{1}{4}g^{\mu\rho}g^{\nu\sigma}
[\nabla_\sigma(\nabla\chi)^2\delta g_{\nu\rho}-\nabla_\rho(\nabla\chi)^2\delta g_{\nu\sigma}]\nn\\
&&+\ft12 g^{\mu\lambda}\nabla^\rho\chi\nabla^\sigma\chi\nabla_\rho\delta g_{\sigma\lambda}-\ft12\nabla_\rho(\nabla^\mu\chi\nabla^\sigma\chi)
g^{\rho\lambda}\delta g_{\sigma\lambda}\nn\\
&&-\ft{1}{4}g^{\mu\lambda}\nabla^\rho\chi\nabla^\sigma\chi\nabla_\lambda\delta g_{\rho\sigma}+\ft{1}{4}\nabla_\lambda(\nabla^\rho\chi\nabla^\sigma\chi)
g^{\lambda\mu}\delta g_{\rho\sigma}\nn\\
&&-\ft{1}{4}g^{\rho\lambda}\nabla^\mu\chi\nabla^\sigma\chi\nabla_\sigma\delta g_{\rho\lambda}+\ft{1}{4}\nabla_\sigma(\nabla^\sigma\chi\nabla^\mu\chi)
g^{\rho\lambda}\delta g_{\rho\lambda}\,.
\eea
Following the Wald procedure, we can now define a 1-form $J_\1=J_\mu dx^\mu$ and its Hodge dual
\be
\Theta_{\sst{(n-1)}}=(-1)^{n+1}{*J_{\1}}\,.
\ee

   We now specialise to a variation that is induced by an infinitesimal
diffeomorphism $\delta x^\mu=\xi^\mu$.  One can show that
\be
J_{\sst{(n-1)}}\equiv \Theta_{\sst{(n-1)}} - i_{\xi} {*L_0} = -
d{*J_\2}\,,
\ee
after making use of the equations of motion.  Here $i_\xi$
denotes a contraction of $\xi^\mu$ on the first index of the
$n$-form ${*L_0}$. One can thus define an $(n-2)$-form
$Q_{\sst{(n-2)}}\equiv {*J_\2}$, such
that $J_{\sst{(n-1)}}=dQ_{\sst{(n-2)}}$.  Note that we use the subscript
notation ``$(p)$'' to denote a $p$-form. To make contact with the first
law of black hole thermodynamics, we take $\xi^\mu$ to be the time-like
Killing vector that is null on the horizon.  Wald shows that the
variation of the Hamiltonian with respect to the integration constants of
a specific solution is given by
\be
\delta {\cal H}=\fft1{16\pi}\, \delta \int_c J_{\sst{(n-1)}} -
\fft1{16\pi}\, \int_c d(i_\xi \Theta_{\sst{(n-1)}}) =\fft{1}{16\pi}\int
_{\Sigma^{(n-2)}} \Big(\delta Q_{\sst{(n-2)}} -
i_\xi \Theta_{\sst{(n-1)}}\Big)\,,\label{deltaH}
\ee
where $c$ denotes a Cauchy surface and $\Sigma^{(n-2)}$ is its boundary,
which has two components, one at infinity and one on the horizon.  Thus
according to the Wald formalism, the first law of black hole thermodynamics
is a consequence of
\be
\delta {\cal H}_{\infty} =\delta {\cal H}_{+}\,.\label{waldidentity}
\ee

For the Horndeski gravity considered in this paper, we find
\bea
J_{\alpha_1\cdots \alpha_{n-1}}
&=&E.O.M+2\epsilon_{\alpha_1\cdots \alpha_{n-1}\,\mu}
\nabla_\nu\left\{\kappa\nabla^{[\nu}\xi^{\mu]}-
\ft{1}{4}\gamma(\nabla\chi)^2\nabla^{[\nu}\xi^{\mu]}+
\ft12\gamma\nabla^{[\nu}(\nabla\chi)^2\xi^{\mu]}\right.\cr
&&\left.+\ft12\gamma\nabla^\sigma\chi\nabla^{[\nu}\chi\nabla_\sigma\xi^{\mu]}
-\ft12\gamma\nabla_\sigma(\nabla^\sigma\chi\nabla^{[\nu}\chi)\xi^{\mu]}
-\ft12\gamma\nabla^{[\nu}(\nabla^{\mu]}\chi\nabla^\sigma\chi)\xi_\sigma\right\}\,,\cr
Q_{\alpha_1\cdots\alpha_{n-2}} &=&
\epsilon_{\alpha_1\cdots \alpha_{n-2}\, \mu\nu}\Big\{
\fft{\partial L}{\partial R_{\mu\nu\rho\sigma}} \nabla_\rho\xi_\sigma
-2\xi_{[\sigma} \nabla_{\rho]} \Big(\fft{\partial L}{\partial R_{\mu\nu\rho\sigma}}\Big) \Big\}\cr
&=&\epsilon_{i_1\cdots i_{n-2}\mu\nu}\left\{\kappa\nabla^\mu\xi^\nu
-\ft{1}{4}\gamma(\nabla\chi)^2\nabla^\mu\xi^\nu +\ft12\gamma\nabla^\sigma\chi\nabla^\mu\chi\nabla_\sigma\xi^\nu\right.\cr
&&\left. +\ft12\gamma\big(\nabla^\mu(\partial\chi)^2\big) \xi^\nu -
\ft12\gamma\nabla_\sigma(\nabla^\sigma\chi\nabla^\mu\chi)\xi^\nu-
\ft12\gamma\nabla^\mu(\nabla^\nu\chi\nabla^\sigma\chi)\xi_\sigma\right\}\,,\cr
(i_\xi \Theta)_{\alpha_1\cdots\alpha_{n-2}} &=&
\epsilon_{\alpha_1\cdots \alpha_{n-2}\,\mu\lambda} \Big( 2 \fft{\partial L}{\partial R_{\rho\sigma\mu\nu}} \nabla_\sigma \delta g_{\rho \nu} -2 \nabla_\nu \fft{\partial L}{\partial R_{\rho\mu\nu\sigma} }\delta g_{\rho\sigma} + \fft{\partial L}{\partial(\nabla_\mu \chi)} \delta \chi \Big) \xi^\lambda\,.\label{JQ}
\eea
To specialise to our static black hole ansatz (\ref{bhansatz}), the
result for the Lagrangian with $\gamma=0$ is well established
(see, for example, \cite{Liu:2013gja,Lu:2014maa}), and is given by
\bea
Q_{\kappa\,, \alpha} &=& r^{n-2}  \kappa \sqrt{\frac{f}{h}} h'\, \Omega_{(n-2)}\,, \cr
i_\xi \Theta_{\kappa\,, \alpha} &=&  -  r^{n-2}  \sqrt{\frac{h}{f}} \Big( \kappa(-\frac{f}{h} \delta h' + \frac{f h'}{2 h^2} \delta  h - \frac{h'}{2 h} \delta f - \frac{n-2}{r} \delta f ) - \alpha f \chi' \delta \chi \Big) \, \Omega_{(n-2)}\,, \cr
(\delta Q -i_\xi \Theta)_{\kappa,\alpha}& =& -  r^{n-2}  \sqrt{\frac{h}{f}} \Big(\kappa \frac{n-2}{r}\delta f + \alpha f \chi' \,\delta \chi\Big) \,
\Omega_{(n-2)}\,,\label{waldmin}
\eea
We find that the contributions associated with the $\gamma$ term in
the action are given by
\bea
Q_\gamma &=&  - \ft{1}2(n-2) \gamma\, r^{n-3}  \sqrt{\frac{h}{f}} f^2 \chi'^2 \Omega_{(n-2)}\,,\cr
i_\xi \Theta_\gamma &=& - \ft{1}2(n-2) \gamma\, r^{n-3}  \sqrt{\frac{h}{f}} f^2 \big( \chi'^2 \fft{\delta h}{2 h}  + (\fft{n-3}{r} (1-\fft{\epsilon}{f}) +  \fft{h'}{h}) \chi'\delta \chi\big)\Omega_{(n-2)}\,,\cr
(\delta Q -i_\xi \Theta)_\gamma &=&  \ft{1}2(n-2)\gamma\, r^{n-3} \sqrt{\frac{h}{f}} f^2 \Big(-  \ft32 \chi'^2 \fft{\delta f}{f} - \delta (\chi'^2) \cr
&&\qquad\qquad\qquad\qquad\qquad+  (\fft{n-3}{r} (1- \fft{\epsilon}{f})+  \fft{h'}{h}) \chi'\delta \chi\Big) \, \Omega_{(n-2)}\,,\label{waldgamma}
\eea

We now apply the Wald formalism to the black hole solutions.  First,
we note that as a consequence of equation (\ref{specialeom}), when we add
the contributions in (\ref{waldmin}) and (\ref{waldgamma}) the
$\chi'\delta\chi$ in the total expression cancel, giving the
result
\be
\delta Q - i_\xi \Theta = -(n-2) r^{n-3}  \sqrt{\frac{h}{f}} \Big[
  (\kappa -\ft34 \gamma f {\chi'}^2)\, \delta f +
  \gamma f^2\, \delta({\chi'}^2)
  \Big]\,.\label{fullvar1}
\ee
In fact, as can be seen from the expression for ${\chi'}^2$ in
for the black hole solutions in (\ref{gensol}),
we have $\delta(f {\chi'}^2)=0$, and so (\ref{fullvar1}) can be further
simplified, to give
\be
\delta Q - i_\xi \Theta = -(n-2) r^{n-3}  \sqrt{\frac{h}{f}} \,
  \Big(\kappa +\ft14 \gamma f {\chi'}^2\Big)\, \delta f \,.\label{fullvar2}
\ee

We first consider the simpler case of the $\epsilon=0$ AdS planar black holes,
for which $f \chi'^2=\beta$.  We find
\bea
\delta {\cal H}_\infty &=& \fft{(n-2)\kappa}{16\pi} \Big(1 + \fft{\beta \gamma}{4\kappa}\Big)\, \delta\mu\,,\cr
\delta {\cal H}_+ &=& \fft{(n-1)(n-2)g^2\kappa}{16\pi}
\Big(1 + \fft{\beta \gamma}{4\kappa}\Big) r_0^{n-2}\, \delta r_0\,.
\eea
Thus we see indeed that $\delta H_{\infty}=\delta H_+$, since $\mu=g^2 r_0^{n-1}$.  This implies that that we can define the mass and entropy as
\be
M=\fft{(n-2)\kappa}{16\pi} \Big(1 + \fft{\beta \gamma}{4\kappa}\Big)\,\mu\,,
\qquad S=\ft14 \kappa \Big(1 + \fft{\beta \gamma}{4\kappa}\Big) r_0^{n-2}\,,
\label{entropy3}
\ee
such that
\be
\delta H_\infty = \delta M\,,\qquad \delta H_+= T\delta S\,.
\ee
The first law of black hole thermodynamics (\ref{firstlaw}) then follows
straightforwardly from the Wald identity (\ref{waldidentity}).  However
the factor $1 + \beta\gamma/(4\kappa)$ in both the
entropy and the mass disagrees with the results in (\ref{S0}) and
(\ref{wfmass1}) that we obtained in section \ref{waldentropysec}
from a direct application of the Wald entropy formula (\ref{waldentropy})
and the integration of the first law $dM=TdS$.

   The case of the spherically-symmetric black holes with
$(\epsilon=1)$ is more complicated.  We find that $\delta {\cal H}$
evaluated on the horizon takes the general form
\be
\delta {\cal H}_+  = \fft{(n-2) \, \omega_{n-2}\, T}{64}
\, (16\kappa  +  \gamma f_1\tilde \chi_1^2) \,
r_0^{n-3}\delta r_0\,,
\ee
where $f_1$ and $\tilde \chi_1$ are coefficients in the near-horizon
expansions defined in (\ref{hfexpan}) and (\ref{phiexpan}).  For our
specific $\epsilon=1$ solution, we have
\bea
f_1 &=& (n-1) g^2 r_0 + \fft{n-3}{r_0}\,,\qquad
\tilde \chi_1=\fft{2\sqrt{(n-1)\beta} g r_0^{\fft32}}{(n-1) g^2r_0^2 + n-3}\,,\cr
h_1 &=& \fft{\Big((n-1)(4\kappa + \beta\gamma) g^2 r_0^2 + 4(n-3)\kappa
\Big)^2}{(4\kappa + \beta\gamma)\big((n-2)g^2r_0^2 + n-3\big)r_0}\,.
\eea
Thus if we define $\delta {\cal H}_+=T dS$, with $T$ given in (\ref{temp2}), we find that the entropy is given by
\bea
S &=& \omega_{n-2}\Big[\ft14 \kappa r_0^{n-2}+ \fft{(n-1)(n-2)\beta\gamma g^2 r_0^n}{
16n(n+2)(n-3)^2} \Big((n+2)(n-3) \cr
&&\qquad\qquad\qquad- n(n-1) g^2 r_0^2\, {}_2F_1[1,\ft12(n+2);
\ft12 (n+4); -\ft{n-1}{n-3} g^2 r_0^2]\Big)\Big]\,.
\eea
Note that the first term inside the square brackets gives precisely the
result we saw earlier (\ref{S0}) for Wald entropy $S_{\sst W}$, derived using
the formula (\ref{waldentropy}).  The remaining contribution in the
square brackets is proportional to $\gamma$, the coefficient of the
Horndeski term in the action (\ref{action}).

To derive the first law, we evaluate the $\delta {\cal H}$ at
asymptotic infinity, and we find
\be
\delta H_\infty = \fft{(n-2)\kappa\, \omega_{n-2}}{16\pi}
\Big(1 + \fft{\beta \gamma}{4\kappa}\Big)\, \delta\mu\,,
\ee
This implies that the mass is given by
\be
M=\fft{(n-2)\kappa\, \omega_{n-2}}{16\pi}
 \Big(1 + \fft{\beta \gamma}{4\kappa}\Big)\mu\,.\label{genmass}
\ee
This turns out to be the exactly the same form as that in the $\epsilon=0$
AdS planar black hole.  It is now straightforward to verify that the
first law (\ref{firstlaw}) is indeed satisfied. Note that $\chi_0$, being
a constant shift integration constant of $\chi$, plays no role in the
first law.

It is worth commenting that for the $\epsilon=0$ solutions, the masses
we obtained in (\ref{wfmass1}) and in (\ref{entropy3})
by the two different methods are both proportional to $\mu$.
The only difference is in the constant prefactor coefficient.
This on its own makes it difficult to judge which is the more reasonable
result.  However, when $\epsilon=1$, the difference becomes more striking.
The result (\ref{genmass}) from the detailed Wald procedure that we
presented in this paper
is seemingly more plausible, for two reasons.  Firstly,  the mass is
simply proportional to the parameter $\mu$, instead of being a
convoluted transcendental function of $\mu$. Secondly, the mass
dependence on $\mu$ is the same for both the $\epsilon=0$ and $\epsilon=1$
solution.  In solutions with no additional scalar hair, and since
the $\epsilon=0$ solution can be obtained as a scaling limit of the
$\epsilon=1$ solution, this conclusion would seem to be reasonable.

\subsection{Further comments on the entropy from Wald formalism}

Having derived the first law of thermodynamics and also the entropy
in section \ref{waldformalism}, using
the general Wald formalism, we now examine the somewhat unusual
features of the
black holes in Horndeski gravity that lead to the
breakdown of the standard Wald entropy formula (\ref{waldentropy}).
It follows from (\ref{JQ}) that for the static ansatz (\ref{bhansatz}) that
\be
Q_{(n-2)} = 2h' \sqrt{\fft{f}{h}}\, S^{\hat 0\hat 1\hat 0\hat 1}_{(n-2)} -
4h T^{0101}{}_{;1} \Omega_{(n-2)}\,,\label{absQ}
\ee
where the hatted indices are tangent-space indices, the semicolon
denotes a covariant derivative and
\be
T^{\mu\nu\rho\sigma}\equiv \fft{\partial L}{\partial
   R_{\mu\nu\rho\sigma}}\,,\qquad
S^{\hat 0\hat 1\hat 0\hat 1}_{(n-2)} =
T^{\hat 0\hat 1\hat 0\hat 1} r^{n-2}
\Omega_{(n-2)}\,.
\ee
Note that 0 is the time direction and 1 is the $r$ direction.  The
expression for $T^{\mu\nu\rho\sigma}$ for the Horndeski gravity
is given by (\ref{dLdR}).
Typically, one evaluates $Q_{(n-2)}$ on the horizon at $r=r_0$,
with $h=h_1 (r-r_0) + \cdots$ and $f=f_1 (r-r_0) + \cdots$, and so
the second term on the right-hand side of (\ref{absQ}) vanishes and hence,
as was observed in \cite{wald1,wald2}, we find
\be
\fft{1}{16\pi} \int_{r=r_0} Q_{(n-2)} = T S_{\sst W}\,,\label{Qres}
\ee
where $S_{\sst W}$ is the standard Wald entropy, given by (\ref{waldentropy}).

Establishing the variational identity (\ref{waldidentity}) is more
subtle, even for the standard case of Einstein gravity.  It requires that
we evaluate $\delta Q$ on the horizon.  Naively, one would simply obtain
$\delta T S_{W} + T \delta S_{\sst W}$ from (\ref{Qres}), and then one would
expect that the $\delta TS_{\sst W}$ term would be cancelled by the
$i_\xi\Theta$ contribution in (\ref{deltaH}), leading to
\be
\delta {\cal H}_+=T \delta S_{\sst W}\,.
\ee
However, in order to evaluate the variation properly, we need to
expand (\ref{Qres}) up to order $(r-r_0)$, since $\delta (r-r_0)=
-\delta r_0$ and so it is non-zero even in the limit when one
sets $r=r_0$ on the horizon.
The net effect is that all the terms in $\delta Q_{(n-2)}$ are cancelled
out by terms in $i_\xi \Theta$, and in fact the $T\delta S$ term arises
from the
remaining terms in $i_\xi \Theta$ alone.

   To be specific, let us examine
$\delta Q - i_\xi \Theta$ for a spherically-symmetric black
hole in pure Einstein gravity coupled to a massless
scalar, as given by (\ref{waldmin}).  If we first perform Taylor expansions
of $Q$ and $i_\xi\Theta$, as given in the first two equations in
(\ref{waldmin}), around the horizon  at $r=r_0$, then indeed the
above statement can be verified.  The final equation in (\ref{waldmin})
gives an alternative but equivalent evaluation with the variation $\delta Q$,
which makes the observation more apparent. We may evaluate $\delta Q$ first,
and then set $r=r_0$.  In this case, the $r^{n-2}$ factor in
$Q_{\kappa,\alpha}$ just depends on the coordinate $r$, and hence is
not varied.  With this procedure, we find that all the terms in
$\delta Q_{\kappa,\alpha}$ are cancelled out by terms in
$i_\xi\Theta_{\kappa,\alpha}$, leading to the third equation of
(\ref{waldmin}).  Thus using this procedure, we find that the
$\delta {\cal H}_+=T\delta S$ term for the usual Einstein gravity arises from
the $(n-2) \delta f/r$ term in $i_\xi\Theta$ in (\ref{waldmin}).  This term
corresponds to
\be
r^{n-2}\sqrt{\fft{h}{f}}\, \fft{2}{rf}\, g_{ij} T^{1i1j}
\delta f\, \Omega_{(n-2)}\,.
\ee
It is rather intriguing how this term is ultimately related to $S_{\sst W}$
which
involves only $T^{0101}$.  Indeed, we see
from (\ref{dLdR}) that in vielbein components,
$T^{\hat 0\hat 1\hat 0\hat 1}= -\ft12\kappa$ and
$T^{\hat 1 \hat i \hat 1\hat j}= \ft12\kappa\,
\delta^{ij}$ for the Horndeski black hole solutions.
In particular, the $\gamma$ term does not contribute in either
case.

   In the black holes of Horndeski gravity there are further subtleties.
Firstly, the $\alpha$ term in $i_\xi \Theta_{\kappa,\alpha}$ in (\ref{waldmin})
does not vanish for these solutions, and can contribute a term to the
entropy that is
not contained in $S_{\sst W}$.  Furthermore, although the second term in
(\ref{absQ})
vanishes on the horizon, its variation does not.  This extra term can be seen
in the form of $Q_\gamma$ in (\ref{waldgamma}).
Thus $(\delta Q -  i_\xi \Theta)_\gamma$ in (\ref{waldgamma}) will give
an additional contribution to the entropy that is over and above that
of the standard Wald contribution $S_{\sst W}$.  Thus we now have
\be
\delta {\cal H}_+=T\delta S\,,\qquad\hbox{with}\qquad S\ne S_{\sst W}\,.
\ee
However,  the Wald identity (\ref{waldidentity}), as we have seen,
continues to hold.
The non-vanishing contributions from both the $\alpha$ and the $\gamma$
terms have the same essential origin, namely that the scalar field
$\chi$ is not regular on the horizon, but rather, it has a branch cut
singularity, as shown in (\ref{phiexpan}).

   One might question whether this is compatible with the interpretation
of the solutions as black holes.   However, as we have remarked in
section \ref{thetheory},
the scalar $\chi$ in Horndeski gravity is like an axion, in the
sense that it enters the theory only through its derivative. In particular,
therefore, it would not be natural to define different
conformally-scaled metric frames (in the manner that one does with
the dilaton in string theory), since that would break the manifest
axionic shift symmetry of $\chi$.  Furthermore, all invariant polynomials
constructed from $\partial_\mu \chi$ with the metric and the Riemann
tensor are regular on the horizon.  For example, $g^{\mu\nu}\del_\mu\chi\,
\del_\nu\chi$ is finite and non-zero on the horizon.  (These
properties can be seen from the fact that
the vielbein components of the gradient of $\chi$ are finite everywhere,
including on the horizon, since one just has $E_{\hat 1}^\mu\,
\del_\mu\chi= \sqrt{f}\, \chi' =
 \sqrt{\beta}\, [1-(n-3)\epsilon/((n-2) g^2 r^2)]^{-1/2}$, with all
other components vanishing, where $E^\mu_{\hat a}$ is the inverse vielbein.)
This supports the idea that these solutions admit a valid
black hole interpretation, but at the price that the Wald entropy formula
(\ref{waldentropy}) no longer provides the complete expression
for the entropy.  However, the identity (\ref{waldidentity}), and hence
the first law of black hole thermodynamics, continue to hold, with the
entropy being derived from the strict application of the Wald formalism.

\subsection{Noether charge and mass of AdS planar black holes}

In the previous subsections, we described two different methods for
calculating the entropy and mass of the Horndeski black holes, one based
on the use of the Wald formula (\ref{waldentropy}) for the entropy, and the
other based on a more detailed consideration of the Wald formalism.
In both these approaches, we did not use independent procedures to
calculate the mass and entropy, but rather, we relied on the use of
the first law of thermodynamics to obtain one from the other.
Since the black-hole solutions are characterised by only one parameter,
there is no non-trivial integrability check, in the sense that the right-hand
side of the first law $dM=TdS$ would be integrable
regardless of whether the expression for the entropy was correct or not.
The fact that the two approaches led to different results calls for
an independent check on the calculation of the mass, or the entropy.
Even though the mass and entropy obtained from the Wald formalism in
section 3.2 seems to be more reasonable, the mass is determined through
an integration of the first law, rather than directly, in this case.
A question one can ask is whether the mass is indeed a conserved quantity.

   For the AdS planar black holes (i.e. the $\epsilon=0$ solutions),
this question can be answered by means of a simple Noether calculation.
For $\epsilon=0$, we rewrite the ansatz as
\be
ds^2 = d\rho^2 - a(\rho)^2dt^2 + b(\rho)^2 d\Omega^2_{\epsilon}\,,\qquad
\chi=\chi(\rho)\,.
\ee
The effective one-dimensional Lagrangian becomes
\bea
L&=& \fft{1}{16\pi} ab^{n-2}\Big(\kappa (R-2\Lambda_0) - \ft12\alpha \chi'^2 + \ft12
\gamma G_{11} \chi'^2\Big)\,,\cr
R &=& -\fft{2a''}{a} - \fft{2(n-2)b''}{b} -
\fft{2(n-2)a'b'}{ab} - \fft{(n-2)(n-3) b'^2}{b^2} + \fft{\epsilon (n-2)(n-3)}{b^2}\,,\cr
G_{11} &=& \fft{(n-2)a'b'}{ab} + \fft{(n-2)(n-3) b'^2}{2 b^2} -\fft{\epsilon (n-2)(n-3)}{2 b^2}\,.
\eea
where a prime denotes a derivative with respect to $\rho$.
The Lagrangian is invariant under the global scaling
\be
a\rightarrow \lambda^{2-n}\,a\,,\qquad b\rightarrow \lambda\, b\, \,.
\ee
This global symmetry yields a conserved Noether charge
\be
{\cal Q}_N=
\fft{1}{16\pi} (n-2) b^{n-3} (b a' - a b') (4\kappa + \gamma \chi'^2)\,.
\ee
In terms of the coordinates of the original ansatz (\ref{bhansatz}),
we have
\be
{\cal Q}_N
=\fft{n-2}{32\pi}r^{n-3} \sqrt{\fft{f}{h}}\, (r h' - 2h) (4\kappa + \gamma f \chi'^2)\,.
\ee
Substituting the AdS planar black hole solution into this Noether charge
formula, we find
\be
{\cal Q}_N
=\fft{(n-1)(n-2)\kappa}{8\pi} \Big(1+ \fft{\beta\gamma}{4\kappa}\Big) \mu=
2(n-1) M\,.
\ee
Thus we see that ${\cal Q}_N$ is the same as the mass obtained from the Wald
formalism in section 3.2, up to some purely numerical constants. This
supports the conclusion that the mass and entropy obtained in section 3.2
are valid, whilst the results in section 3.1 are not.

\subsection{Euclidean action}

   An alternative method that has been used
for calculating thermodynamic quantities for black hole solutions is
by means of the {\it quantum statistical relation}
\be
\Phi_{\rm thermo} \equiv M - T S = I\,  T\,,\label{qsr}
\ee
first proposed for quantum gravity in \cite{Gibbons:1976ue}.
Here $\Phi_{\rm thermo}$ denotes the thermodynamic potential, or the
free energy, and $I$ is the Euclidean action. The regularised Euclidean action
was calculated for the $\epsilon=1$ Horndeski black hole in four
dimensions in \cite{aco}.  We have repeated that calculation,
and obtained the same result (save for an overall factor of 2 discrepancy).
However, the resulting expressions for $M$ and entropy are quite
different from those in sections 3.1 or 3.2, and are given by
\bea
M &=& \ft12\kappa \Big(1 + \fft{\beta\gamma}{4\kappa}\Big) \mu -
\fft{3\beta\gamma g^2 r_0^3 \big(4\kappa (3g^2 r_0^2+1) +
3\beta \gamma g^2 r_0^2\big)}{8(4\kappa + \beta\gamma)
(1 + 3 g^2 r_0^2)\big(4\kappa (3g^2 r_0^2-1) + 3\beta\gamma g^2 r_0^2\big)}
\,,\cr
S&=& \kappa \pi r_0^2 + \fft{3\pi\beta\gamma\, g^2 r_0^4 \,
\big(4\kappa + 3 (4\kappa +\beta\gamma) g^2 r_0^2\big)}{2(1 + 3 g^2 r_0^2)
\big(4\kappa - 3(4\kappa+\beta\gamma) g^2 r_0^2\big)}\,.\label{ep1MS}
\eea
Note that when $\beta=0$, for which the black hole reduces to the
standard Schwarzschild-AdS one, we get $M=\ft12\mu$
and $S=\kappa \pi r_0^2$, as one would expect. It is clear that the
mass suffers from the same shortcoming as the one we obtained from the
Wald entropy formula in (\ref{wfmass2}), in that it becomes a
convoluted transcendental function of
$\mu$ for non-vanishing $\beta$.  (It is a different transcendental function
from the one following from (\ref{wfmass2}), however.)

The calculation for the $\epsilon=0$ AdS planar black holes (\ref{planarso})
is much easier,  and can be straightforwardly carried out
for a general spacetime dimension $n$. The regularised Euclidean action
can be
defined by subtracting the action of the background $\mu=0$
vacuum from the action for the black hole itself, namely
\be
I_{reg} = I_E [g_{\mu\nu}\,, \chi] - I_E [g^\0_{\mu\nu}\,, \chi^\0]\,,
\ee
where $g^\0_{\mu\nu}$ and $\chi^\0$ are the background field obtained by
setting $\mu =0$ in the black hole solution (\ref{planarso}). We find
\be
I_{reg} = - \fft{\kappa}{16(n-1)}\Big(1 -
    \fft{(n-2) \beta \gamma}{4 \kappa}\Big) r_0^{n-2} \,.
\ee
Note that in this calculation,  we have set $\omega_{n-2}=1$, so that the
resulting extensive quantities are densities. Using the quantum statistical
relation (\ref{qsr}) and the thermodynamic  first law (\ref{firstlaw}),
we then find that the free energy, mass, temperature and entropy for the
$\epsilon=0$ black holes are given by
\bea
F &=& -\fft{\kappa\mu}{16\pi}
\Big(1 - \fft{(n-2)\beta\gamma}{4\kappa}\Big)\,,\qquad
M=-(n-2) F\,,\cr
T &=& \fft{g^2(n-1) r_0}{4\pi}\,,\qquad
S=\ft14\kappa r_0^{n-2}  - \ft1{16} (n-2) \beta\gamma r_0^{n-2}\,.
\eea
These expressions also disagree, in this case by constant overall factors,
with the $\epsilon=0$ results obtained in sections 3.1 and 3.2.
Taken in isolation, it would be hard to make any judgment as to
whether these expressions were trustworthy or not.
Interestingly the generalized Smarr relation (\ref{gsmarr}) is also
satisfied. However, the $\epsilon=1$ results (\ref{ep1MS}) for
the mass and the entropy certainly raise questions about the
validity of this calculation using the Euclidean action.

There is another method that has been used in order to obtain a
finite Euclidean action,  by adding a surface term and a counterterm.
Taking $n=4$ dimensions as an example, the whole action is then given by
\be
I = I_{bulk} - 2 I_{\rm GH} -  I_{\rm ct} \,,
\ee
where $I_{\rm GH}$ is the standard Gibbons-Hawking surface term,
and for   $\epsilon=0$, the counterterm is given by
\be
I_{\rm ct} = \kappa \int dx^3 \sqrt{\gamma}  c_1 g \,,
\quad \hbox{with} \quad c_1 = 4 + \fft{\beta \gamma}{\kappa} \,,
\ee
The $\gamma $ in the square root is the determinant of induced metric
$\gamma_{\mu\nu}$. With these combinations, the total action is the same
as the result of regularization.
For $\epsilon = 1$, the counterterm is
\be
I_{\rm ct} = \kappa \int dx^3 \sqrt{\gamma}
( c_1 g + \fft{c_2 R[\gamma]}{g} ) \,, \quad
\hbox{with} \quad
c_1 = 4 + \fft{\beta \gamma}{\kappa} \,, \quad
c_2 = 1 - \fft{\beta \gamma}{4 \kappa}
\ee
and the value of the action has an additional term linear the imaginary-time
period (i.e. inversely proportional to the temperature),
in comparison to that of the regularized calculation above:
\be
I_{\rm renorm} = I_{\rm reg} + \fft{\sqrt 3 \pi \beta^2 \gamma^2}{
12 g (4 \kappa + \beta \gamma) } \fft{\epsilon}{T}\,.
\ee
The effect on the thermodynamics is that the entropy is unchanged,
but the mass acquires an additive contribution in the
spherically-symmetric $\epsilon=1$ solutions, independent of the
parameter in the solutions. This is not surprising, since when $\epsilon=1$,
the $\mu=0$ solution is not vacuum AdS spacetime, but instead
a smooth soliton, which has a constant mass.  In the earlier regularisation
by subtracting the background, this constant energy was subtracted out.

   The question remains as to how one might reconcile the results
for the entropy and the mass, as calculated from the regularised Euclidean
action, with our previous, and different, results obtained using the
Wald formalism.  We do not have a definitive resolution to this puzzle,
other than to suggest that because of the rather unusual features of
the black-hole solutions in Horndeski gravity, it may be that the naive
application of a subtraction procedure to obtain a regularised
Euclidean action may be inherently ambiguous.  In a somewhat related
context, it was found in \cite{chlupo} that attempts to employ the
Abbott-Deser method \cite{abodes}
to calculate the mass of asymptotically-AdS
black holes foundered on ambiguities in the subtraction procedure in
some cases, for solutions in gauged supergravities where scalar fields
were involved.  In the absence of a rigorous derivation of a valid
subtraction scheme for the calculation of the Euclidean action, it
seems that one could engineer different schemes that gave different
results, with no guide as to which result should be regarded as the
correct one.

\section{Viscosity/Entropy Ratio}

One of the motivations for this paper was to study the
viscosity/entropy ratio in Horndeski gravity.  Having obtained a
formula for the entropy of the black holes, we are now in a
position to proceed.  To calculate the shear viscosity of the
boundary field theory, we consider a transverse and traceless
perturbation of the AdS planar black hole, namely
\be
ds^2 =-f dt^2 + \fft{dr^2}{f} + r^2 \big(dx_i dx_i +
  2 \Psi(r,t) dx_1 dx_2\big)\,,
\ee
where the background solution is given by
(\ref{alphabeta}), (\ref{planarso}) and (\ref{phiso1}).  We find that the
mode $\Psi(r,t)$ satisfies the linearised equation
\bea
&&r\,(4 \kappa +\beta \gamma) (g^2 r^{n-1} -\mu)^2 \, \Psi''
 + (4 \kappa +\beta \gamma)  (g^2 r^{n-1} -\mu)
(n g^2 r^{n-1} -\mu)\, \Psi'\cr
&&\qquad\qquad - r^{2n-5}\, (4 \kappa -\beta \gamma)\, \ddot\Psi =0\,.
\label{perteom}
\eea
For an infalling wave which is purely ingoing at the horizon, the solution
for a wave with low frequency $\omega$ is given by
\bea
\Psi &=& e^{-{\rm i} \omega t} \psi(r)\,,\qquad \psi(r)=
\exp(-{\rm i} \omega K \log\fft{f(r)}{g^2 r^2}) + {\cal O}(\omega^2)\,,\cr
K &=&\fft{1}{4\pi T} \sqrt{\fft{4 \kappa -\beta \gamma}{4 \kappa +\beta \gamma}}\,. \label{pereq}
\eea
Note that the constant parameter $K$ is determined by the horizon boundary condition.
The overall integration constant is fixed so that $\Psi$ is unimodular
asymptotically, as $r\rightarrow \infty$.

In order to study the boundary field theory using the AdS/CFT correspondence,
we substitute the ansatz with the linearised perturbation into the action.
The quadratic terms in the Lagrangian, after removing the
second-derivative contributions using the Gibbons-Hawking term, can
be written as
\be
{\cal L}_2 = P_1\, {\Psi'}^2 + P_2\, \Psi\,\Psi' + P_3\, \Psi^2
+ P_4\, \dot\Psi^2 \,,
\ee
with
\bea
P_1 &=& - \fft18 (4 \kappa +\beta \gamma) \, (g^2 r^{n-1}-\mu)r\,, \quad P_2 = \fft12 g^2 r^{n-1} [4 \kappa - (n-2) \beta \gamma] - \mu (2 \kappa - \fft{n-3}{4} \beta \gamma) \,,  \cr
P_3 &=& \fft{n-1}{4} g^2 r^{n-2} [4 \kappa - (n-2) \beta \gamma] \,, \quad P_4 = \fft{r^{2n - 5} (4 \kappa - \beta \gamma)}{8 (g^2 r^{n-1} - \mu)}
\eea
Note that $P_3 = \ft12 P_2'$.  We then find that the terms quadratic in
$\Psi$ in
the Lagrangian are given by
\be
{\cal L}_2 = \fft{d}{dr}\, (P_1 \Psi\,\Psi' + \ft12 P_2\Psi^2)
  +\fft{d}{dt} (P_4\, \Psi\, \dot\Psi) -
   \Psi\, \Big[ P_1\, \Psi'' + P_1'\, \Psi' + P_4\, \ddot{\Psi} \Big]\,.
\ee
The last term , enclosed in square brackets, vanishes by virtue of
the linearised perturbation equation (\ref{perteom}), and so the
quadratic Lagrangian is a total derivative.  The viscosity is determined
from the $P_1 \Psi \Psi'$ term,
following the procedure described in \cite{shenker,sonsta}.  Using
this, we find that the viscosity is given by
\be
\eta= \fft{\kappa (n-1)\mu}{64\pi^2 T}
\sqrt{1 - \fft{\beta^2 \gamma^2}{16\kappa^2}}\,.
\label{eta0}
\ee

   We have, for the planar black holes,
\be
\mu=g^2 r_0^{n-1}\,,\qquad T=\fft{(n-1)g^2 r_0}{4\pi}\,,
\ee
and the entropy that we derived in section 3.2 using the
Wald formalism is given by
\be
S = \fft14 \kappa \Big(1+\fft{\beta \gamma}{4 \kappa}\Big) r_0^{n-2} \,.
\ee
We therefore find that the viscosity/entropy ratio  is given by
\be
\fft{\eta}{S} = \fft{1}{4\pi} \sqrt{ \fft{4\kappa -\beta \gamma}{
 4\kappa + \beta \gamma}}\label{eta1}
\ee
for the Horndeski black holes.\footnote{Intriguingly, although the ratio
is calculated for the AdS planar black hole ($\epsilon=0$), the
same ratio $(4\kappa -\beta \gamma)/(4\kappa + \beta \gamma)$
appears in the sub-leading constant term in the large-$r$
expansion of $h=-g_{tt}$ given in (\ref{larger}), but only for
the spherically-symmetric
($\epsilon=1$) solutions (it vanishes for the $\epsilon=0$ solutions).}
Note that $\kappa$ and $\beta$ are both positive. For reality, we must have
\be
-\fft{4\kappa}{\beta} < \gamma <\fft{4\kappa}{\beta}\,.
\ee
When $\beta = 0$, which turns off the scalar field, the ratio goes back to
the universal value of $1/(4 \pi)$.  When $\gamma > 0$, the ratio is less
than $1/(4\pi)$ and hence the bound is violated. For $\gamma<0$, the
ratio is greater than $1/(4\pi)$.

Finally, we note that in terms of the original parameters of the
theory (\ref{action}), the viscosity/entropy ratio is given by
\be
\fft{\eta}{S} = \fft{1}{4\pi}\,\sqrt{\fft{3\alpha + \gamma \Lambda_0}{\alpha - \gamma \Lambda_0}}\,.
\ee
Interestingly, the ratio is independent of the parameter $\kappa$.

\section{Conclusion}

Motivated by applications for the AdS/CFT correspondence, we studied
the black holes in a theory of Einstein gravity coupled
to a scalar field, including a non-minimal Horndeski term where the
gradient of the scalar couples to the Einstein tensor. There are
two types of static black holes in this Horndeski gravity.  One
of these is the
usual Schwarzschild-AdS black hole, for which the scalar field is constant.
Our focus is on the other non-trivial one-parameter family of
static black holes, for which the scalar depends non-trivially on the
radial coordinate.  Although the scalar has a branch-cut singularity on
the horizon, it is axion-like and enters the theory only through a
derivative. Furthermore, in an orthonormal frame, $\partial_a\chi$ is
regular everywhere, both on and outside the horizon, and all invariants
involving the scalar field are finite everywhere.
We also demonstrated the uniqueness of these static black hole solutions
in the theory.

We studied the thermodynamics of the black holes and found three surprises.
The first is that the standard Wald entropy formula (\ref{waldentropy})
does not give the complete expression for the entropy of these black holes.
This can be attributed to the fact that the derivation of the
Wald entropy (\ref{waldentropy}) requires that the scalar be regular on
the horizon.  In fact, the branch cut singularity of the scalar
on the horizon implies that there is an extra contribution to the entropy.
We studied the Wald formalism in detail,  and exhibited the new
contribution explicitly.  It turns out that the Wald identity
(\ref{waldidentity}) continues to hold for these black holes, and
so does the first law of black hole thermodynamics.  The entropy,
however, is no longer given by (\ref{waldentropy}), but can be determined
from the implementation of the Wald procedure.  We further established,
using a simple construction of the Noether charge derivable from
the scaling symmetry of the
planar black holes, that the mass of the AdS planar black hole,
as we derived from the Wald procedure, is indeed a conserved quantity.

The second surprise concerns the use of the
quantum statistical relation $E-TS =T I$ to calculate the thermodynamic
parameters of the black hole solutions.  In order to apply this method,
it is necessary to calculate the Euclidean action $I$ of the
black hole solution.  The problem is that a direct integration of the
Euclideanised action yields a result that diverges at the upper end of
the radial integration, and so it is
necessary to adopt some regularisation procedure.  We tried
to apply two different such procedures. The first involved subtracting
the diverging contribution of a background where the mass is set to zero
from the diverging contribution from the black hole with non-zero mass.
The other procedure involved adding a
boundary counterterm.  The two methods gave the same results for the
mass and the entropy, but these results differed from those that
we obtained by using the Wald formalism.
The origin of this mismatch is not clear to us; it may be related to
intrinsic ambiguities in the subtraction schemes that we used in order to
 regularise the divergences.  Such ambiguities are possibly more
likely in a theory such as Horndeski gravity, with its somewhat
unusual features, and so regularisation schemes for calculating the
Euclidean action that usually work in less exacting situations may need
to be scrutinised more carefully here.

The third surprise concerns the results in section 4 for the viscosity/entropy
ratio.  In wide classes of conventional theories with no
higher-derivative terms in the Lagrangian, one finds a rather
universal result that $\eta/S=1/(4\pi)$.   Counter-examples to the
universality of the ratio have been found, but for isotropic
situations such as we have considered they are always associated
with higher-derivative gravities, such as Gauss-Bonnet or more
general Lovelock gravities.  As far as we are aware, our findings for
the black holes in the Horndeski theory we studied in this paper
provide the first
example of the violation of the $\eta/S=1/(4\pi)$ result in a theory
whose Lagrangian is at most linear in curvature tensor.

  A word of caution about the use of the Wald formalism to calculate
the entropy is perhaps appropriate here.  If we consider Einstein-Maxwell
theory as an example, the first law $dM=T dS + \Phi dQ$ for
Reissner-Nordstr\"om black holes can be derived from the Wald formalism
by calculating $\delta{\cal H}_\infty$ and $\delta{\cal H}_+$, and
using the fact that $\delta{\cal H}_\infty = \delta{\cal H}_+$.  The
$\Phi dQ$ contribution can either enter in $\delta{\cal H}_+$ alone,
if one uses the gauge where the potential vanishes at infinity, or
in $\delta{\cal H}_\infty$ alone, if one uses the gauge where the
potential vanishes on the horizon, or else in both $\delta{\cal H}_\infty$
and $\delta{\cal H}_+$, if one uses some intermediate gauge where the
potential vanishes neither at infinity nor on the horizon.  In the first
law, only the potential difference $\Phi\equiv\Phi_+ -\Phi_\infty$
contributes. If the gauge where the potential vanishes on the horizon
is chosen, then $\delta{\cal H}_+=T\delta S$ and so $\delta{\cal H}_+/T$
is an exact differential, which can be integrated to give the entropy, while
$\delta{\cal H}_\infty = dM +\Phi_\infty\, dQ$, and is not exact.  In
the gauge where the potential instead vanishes at infinity,
$\delta{\cal H}_\infty = dM$, which is an exact differential, while
$\delta{\cal H}_+=T dS + \Phi_+\, dQ$, and so $\delta{\cal H}_+/T$ is
not exact.

   More complicated situations were encountered recently
where asymptotically-AdS dyonically charged black holes were
constructed in a four-dimensional gauged supergravity involving a scalar
and a Maxwell field \cite{Lu:2013ura,Chow:2013gba}. It was found that
$\delta{\cal H}_\infty$ was non-exact, and hence non-integrable,
even when a gauge where the
electric and magnetic potentials vanished at infinity was chosen,
because of a varying contribution from the asymptotic coefficients in
the large-distance expansion of the scalar field. The first law of
black hole (thermo)dynamics, involving the
scalar contribution, could nevertheless be derived using the strict
Wald formalism \cite{Lu:2013ura}.  The results were later generalised to
black holes in general Einstein-scalar theories\cite{Liu:2013gja,Lu:2014maa},
Einstein-Proca theories \cite{Liu:2014tra}, and gravity extended with
quadratic curvature invariants \cite{Fan:2014ala}.

  Analogous issues could in principle arise when considering
$\delta{\cal H}_+$:
it is commonly the case that  $\delta {\cal H}_+$ on
the horizon can be expressed as $T\delta S$.  In a theory such as
Einstein-Maxwell, this is a gauge-dependent property as we discussed
above, and  in
order to have $\delta {\cal H}_+/T$ be an exact differential in this case
one would need to work in the gauge where the electric potential vanished on
the horizon.  In most theories that have been studied,
the entropy is simply given by $S_W$ defined by
the Wald entropy formula (\ref{waldentropy}).
The widespread validity of the Wald entropy formula is related to the
fact that typically, matter fields vanish on the horizon of
a black hole (and Maxwell potentials can be set to zero by means of
appropriate gauge choices).  In the Horndeski gravity considered in this
paper, however, the axion-like scalar $\chi$ has an unusual behaviour near
the horizon and near infinity, and indeed we have already seen
that $\delta {\cal H}_+\ne T\delta S_W$. We nevertheless assumed
that it was still the case that $\delta {\cal H}_+=T\delta S$, i.e.
that  $\delta {\cal H}_+/T$ could be integrated to define an entropy
function.  That $\delta {\cal H}_+/T$ is integrable is guaranteed in the
one-parameter family of solutions considered in this paper, since all
1-forms in one dimension are exact. In a multiple-parameter black hole
solution, however, there does not appear to be any guarantee, {\it a priori},
that $\delta {\cal H}_+/T$ must be a total differential in a theory
such as Horndeski gravity.  The non-integrability of the sort that
occurs in $\delta {\cal H}_\infty$ in the dyonic asymptotically-AdS
black holes we discussed above might also, in principle, occur
for $\delta {\cal H}_+/T$ on the horizon, if not all the fields are
strictly vanishing on the horizon.  It would be interesting to
study this further in more general solutions in theories such as
Horndeski gravities.

The findings in this paper
indicate that Horndeski gravity, and its black hole solutions
in particular, deserve further investigation both in their own right,
and also in the context of the AdS/CFT correspondence.

\section*{Acknowledgements}

We are grateful to Sera Cremonini for helpful discussions.
H-S.L.~is supported in part by NSFC grants 11305140, 11375153 and 11475148,
SFZJED grant Y201329687 and CSC scholarship No. 201408330017. C.N.P.~is supported in part by DOE grant DE-FG02-13ER42020. The work of X-H.Feng and H.L.~are supported in part by NSFC grants NO. 11175269, NO. 11475024 and NO. 11235003.


\begin{thebibliography}{99}

\bibitem{adscft1} J.M. Maldacena,
{\it The large N limit of superconformal field theories and supergravity},
Adv.\ Theor.\ Math.\ Phys.\  {\bf 2}, 231 (1998),
hep-th/9711200.

\bibitem{adscft2} S.S. Gubser, I.R. Klebanov and A.M. Polyakov,
{\it Gauge theory correlators from non-critical string theory},
Phys.\ Lett.\  {\bf B428}, 105 (1998),
hep-th/9802109.

\bibitem{adscft3} E. Witten,
{\it Anti-de Sitter space and holography},
Adv. Theor. Math. Phys. {\bf 2}, 253 (1998), hep-th/9802150.

\bibitem{adscft4}
  O. Aharony, S.S. Gubser, J.M. Maldacena, H. Ooguri and Y. Oz,
{\it Large $N$ field theories, string theory and gravity,}
  Phys.\ Rept.\  {\bf 323}, 183 (2000)
  [hep-th/9905111].

\bibitem{Policastro:2001yc}
  G. Policastro, D.T. Son and A.O. Starinets,
{\it The shear viscosity of strongly coupled ${\cal N}=4$
supersymmetric Yang-Mills plasma,}
  Phys.\ Rev.\ Lett.\  {\bf 87}, 081601 (2001),
 hep-th/0104066.

\bibitem{sonsta} D.T. Son and A.O. Starinets,
{\it Minkowski space correlators in AdS/CFT correspondence:
Recipe and applications},
JHEP {\bf 0209}, 042 (2002), hep-th/0205051.

\bibitem{KSS} P. Kovtun, D.T. Son and A.O. Starinets,
{\it Holography and hydrodynamics: Diffusion on stretched horizons},
JHEP {\bf 0310}, 064 (2003), hep-th/0309213.

\bibitem{KSS0} P. Kovtun, D.T. Son and A.O. Starinets,
{\it Viscosity in strongly interacting quantum field theories from black
hole physics},  Phys.\ Rev.\ Lett.\  {\bf 94}, 111601 (2005),
hep-th/0405231.

\bibitem{Iqbal:2008by}
  N. Iqbal and H. Liu,
{\it Universality of the hydrodynamic limit in AdS/CFT and
the membrane paradigm,}
  Phys.\ Rev.\ D {\bf 79}, 025023 (2009),
  arXiv:0809.3808 [hep-th].

\bibitem{Cai:2008ph}
  R.G. Cai, Z.Y. Nie and Y.W. Sun,
{\it Shear viscosity from effective couplings of gravitons,}
  Phys.\ Rev.\ D {\bf 78}, 126007 (2008),
  arXiv:0811.1665 [hep-th].

\bibitem{Cai:2009zv}
  R.G. Cai, Z.Y. Nie, N. Ohta and Y.W. Sun,
{\it Shear viscosity from Gauss-Bonnet gravity with a dilaton coupling,}
  Phys.\ Rev.\ D {\bf 79}, 066004 (2009),
 arXiv:0901.1421 [hep-th].

\bibitem{Brustein:2007jj}
  R. Brustein, D. Gorbonos and M. Hadad,
{\it Wald's entropy is equal to a quarter of the horizon area in units of the effective gravitational coupling,}
  Phys.\ Rev.\ D {\bf 79}, 044025 (2009),
  arXiv:0712.3206 [hep-th].

\bibitem{Liu:2015tqa}
  H.S. Liu, H. L\"u and C.N. Pope,
{\it Generalised Smarr formula and the viscosity bound for
Einstein-Maxwell-Dilaton black holes,} Phys.\ Rev.\ D {\bf 92},
no. 6, 064014 (2015), arXiv:1507.02294 [hep-th].

\bibitem{Buchel:2003tz}
  A. Buchel and J.T. Liu,
{\it Universality of the shear viscosity in supergravity,}
  Phys.\ Rev.\ Lett.\  {\bf 93}, 090602 (2004),
  hep-th/0311175.

\bibitem{Buchel:2004qq}
  A.~Buchel,
{\it On universality of stress-energy tensor correlation functions in
supergravity,}
  Phys.\ Lett.\ B {\bf 609}, 392 (2005),
  hep-th/0408095.

\bibitem{Benincasa:2006fu}
  P.~Benincasa, A.~Buchel and R.~Naryshkin,
{\it The shear viscosity of gauge theory plasma with chemical potentials,}
  Phys.\ Lett.\ B {\bf 645}, 309 (2007),
  hep-th/0610145.

\bibitem{Landsteiner:2007bd}
  K.~Landsteiner and J.~Mas,
{\it The shear viscosity of the non-commutative plasma,}
  JHEP {\bf 0707}, 088 (2007),
  arXiv:0706.0411 [hep-th].

\bibitem{Cremonini:2011iq}
  S.~Cremonini,
{\it The shear viscosity to entropy ratio: A status report,}
  Mod.\ Phys.\ Lett.\ B {\bf 25}, 1867 (2011),
  arXiv:1108.0677 [hep-th].

\bibitem{Kats:2007mq}
  Y.~Kats and P.~Petrov,
{\it Effect of curvature squared corrections in AdS on the viscosity of the dual gauge theory,}
  JHEP {\bf 0901}, 044 (2009),
 arXiv:0712.0743 [hep-th].

\bibitem{shenker} M. Brigante, H. Liu, R.C. Myers, S. Shenker and S. Yaida,
{\it Viscosity bound violation in higher derivative gravity},
Phys.\ Rev.\ D {\bf 77}, 126006 (2008), arXiv:0712.0805 [hep-th].

\bibitem{Natsuume:2010ky}
  M. Natsuume and M. Ohta,
{\it The shear viscosity of holographic superfluids,}
  Prog.\ Theor.\ Phys.\  {\bf 124}, 931 (2010),
  arXiv:1008.4142 [hep-th].

\bibitem{Erdmenger:2010xm}
  J. Erdmenger, P. Kerner and H. Zeller,
{\it Non-universal shear viscosity from Einstein gravity,}
  Phys.\ Lett.\ B {\bf 699}, 301 (2011),
arXiv:1011.5912 [hep-th].

\bibitem{Ovdat:2014ipa}
  O. Ovdat and A. Yarom,
{\it A modulated shear to entropy ratio,}
  JHEP {\bf 1411}, 019 (2014),
arXiv:1407.6372 [hep-th].

\bibitem{Ge:2014aza}
  X.H. Ge, Y. Ling, C. Niu and S.J. Sin,
{\it Holographic transports and stability in anisotropic linear axion model,}
  arXiv:1412.8346 [hep-th].

\bibitem{Shu:2009ax}
  F.W. Shu,
{\it The quantum viscosity bound in Lovelock gravity,}
  Phys.\ Lett.\ B {\bf 685}, 325 (2010),
arXiv:0910.0607 [hep-th].

\bibitem{deBoer:2009pn}
  J. de Boer, M. Kulaxizi and A. Parnachev,
{\it AdS$_7$/CFT$_6$, Gauss-Bonnet gravity, and viscosity bound,}
  JHEP {\bf 1003}, 087 (2010),
arXiv:0910.5347 [hep-th].

\bibitem{Camanho:2009vw}
  X.O. Camanho and J.D. Edelstein,
  {\it Causality constraints in AdS/CFT from conformal collider physics
and Gauss-Bonnet gravity,}
  JHEP {\bf 1004}, 007 (2010),
 arXiv:0911.3160 [hep-th].

\bibitem{Brans:1961sx}
  C. Brans and R.H. Dicke,
{\it Mach's principle and a relativistic theory of gravitation,}
  Phys.\ Rev.\  {\bf 124}, 925 (1961).

\bibitem{Horndeski:1974wa}
  G.W. Horndeski,
{\it Second-order scalar-tensor field equations in a four-dimensional space,}
  Int.\ J.\ Theor.\ Phys.\  {\bf 10}, 363 (1974).

\bibitem{Nicolis:2008in}
  A. Nicolis, R. Rattazzi and E. Trincherini,
{\it The Galileon as a local modification of gravity,}
  Phys.\ Rev.\ D {\bf 79}, 064036 (2009),
 arXiv:0811.2197 [hep-th].

\bibitem{Hawking:1974rv}
  S.W. Hawking,
{\it Black hole explosions,}
  Nature {\bf 248}, 30 (1974).

\bibitem{Hawking:1974sw}
  S.W. Hawking,
{\it Particle creation by black holes,}
  Commun.\ Math.\ Phys.\  {\bf 43}, 199 (1975)
  [Erratum-ibid.\  {\bf 46}, 206 (1976)].

\bibitem{wald1}
  R.M. Wald,
{\it Black hole entropy is the Noether charge,}
  Phys. Rev. D {\bf 48}, 3427 (1993), gr-qc/9307038.

\bibitem{wald2}
  V. Iyer and R.M. Wald,
{\it Some properties of Noether charge and a proposal for dynamical
black hole entropy,}
Phys. Rev. D {\bf 50}, 846 (1994), gr-qc/9403028.

\bibitem{aco}
  A. Anabalon, A. Cisterna and J. Oliva,
{\it Asymptotically locally AdS and flat black holes in Horndeski theory,}
  Phys.\ Rev.\ D {\bf 89}, 084050 (2014),
arXiv:1312.3597 [gr-qc].

\bibitem{Rinaldi:2012vy}
  M. Rinaldi,
{\it Black holes with non-minimal derivative coupling,}
  Phys.\ Rev.\ D {\bf 86}, 084048 (2012),
arXiv:1208.0103 [gr-qc].

\bibitem{Babichev:2013cya}
  E. Babichev and C. Charmousis,
{\it Dressing a black hole with a time-dependent Galileon,}
  JHEP {\bf 1408}, 106 (2014),
arXiv:1312.3204 [gr-qc].

\bibitem{Gibbons:1976ue}
  G.W. Gibbons and S.W. Hawking,
{\it Action integrals and partition functions in quantum gravity,}
  Phys.\ Rev.\ D {\bf 15}, 2752 (1977).

\bibitem{Liu:2013gja}
  H.S. Liu and H. L\"u,
{\it Scalar charges in asymptotic AdS geometries,}
  Phys.\ Lett.\ B {\bf 730}, 267 (2014),
arXiv:1401.0010 [hep-th].

\bibitem{Lu:2014maa}
  H. L\"u, C.N. Pope and Q. Wen,
{\it Thermodynamics of AdS black holes in Einstein-scalar gravity,}
  JHEP {\bf 1503}, 165 (2015),
arXiv:1408.1514 [hep-th].

\bibitem{Liu:2014tra}
  H.S. Liu, H. L\"u and C.N. Pope,
{\it Thermodynamics of Einstein-Proca AdS black holes,}
  JHEP {\bf 1406}, 109 (2014),
arXiv:1402.5153 [hep-th].

\bibitem{Fan:2014ixa}
  Z.Y. Fan and H. L\"u,
{\it $SU(2)$-Colored (A)dS black holes in conformal gravity,}
  JHEP {\bf 1502}, 013 (2015),
 arXiv:1411.5372 [hep-th].

\bibitem{Fan:2014ala}
  Z.Y. Fan and H. L\"u,
{\it Thermodynamical first laws of black holes in quadratically-extended
gravities,}
  Phys.\ Rev.\ D {\bf 91}, no. 6, 064009 (2015),
arXiv:1501.00006 [hep-th].

\bibitem{Liu:2014dva}
  H.S. Liu and H. L\"u,
{\it Thermodynamics of Lifshitz black holes,}
  JHEP {\bf 1412}, 071 (2014),
 arXiv:1410.6181 [hep-th].

\bibitem{chlupo} W. Chen, H. L\"u and C.N. Pope,
{\it Mass of rotating black holes in gauged supergravities},
  Phys.\ Rev.\ D {\bf 73}, 104036 (2006),  hep-th/0510081.

\bibitem{abodes} L.F. Abbott and S. Deser,
{\it Stability of gravity with a cosmological constant},
Nucl. Phys. {\bf B195}, 76 (1982).

\bibitem{Lu:2013ura}
  H.~L\"u, Y.~Pang and C.N.~Pope,
  {\it AdS dyonic black hole and its thermodynamics,}
  JHEP {\bf 1311}, 033 (2013)
  [arXiv:1307.6243 [hep-th]].

\bibitem{Chow:2013gba}
  D.D.K.~Chow and G.~Compere,
{\it Dyonic AdS black holes in maximal gauged supergravity,}
  Phys.\ Rev.\ D {\bf 89}, no. 6, 065003 (2014)
  [arXiv:1311.1204 [hep-th]].

\end{thebibliography}
\end{document}